\documentclass[12pt, a4paper, twoside]{fec}
\newcommand{\alfv}{Alfv\'{e}n}

\title{A Systematic Approach to the Linear-Stability Assessment of \alfv{}
Eigenmodes in the Presence of Fusion-Born Alpha Particles for ITER-like
Equilibria}

\author[1]{P. Rodrigues}
\author[1]{A. Figueiredo}
\author[1]{J. Ferreira} 
\author[1]{R. Coelho}
\author[1]{F. Nabais}
\author[1]{D. Borba}
\author[1]{N. F. Loureiro}
\author[2]{H. J. C. Oliver}
\author[3]{S. E. Sharapov}

\affil[1]{Instituto de Plasmas e Fus\~{a}o Nuclear, Instituto Superior
T\'{e}cnico, Universidade de Lisboa, 1049-001 Lisboa, Portugal.}

\affil[2]{HH Wills Physics Laboratory, Univ. of Bristol.}

\affil[3]{EURATOM/CCFE Fusion Association, Culham Science Centre, Abingdon OX14
3DB, UK.}

\doi

\emailID{par@ipfn.ist.utl.pt}

\PaperNumber{TH/P3-25}

\begin{document}

\maketitle

\begin{abstract}
A systematic approach to assess the linear stability of \alfv{} eigenmodes in the
presence of fusion-born alpha particles is described. Because experimental
results for ITER are not available yet, it is not known beforehand which \alfv{}
eigenmodes will interact more intensively with the alpha-particle population.
Therefore, the number of modes that need to be considered in stability
assessments becomes quite large and care must be exercised when choosing the
numerical tools to work with, which must be fast and efficient. In the presented
approach, all possible eigenmodes are first found after intensively scanning a
suitable frequency range. Each solution found is then tested to find if its
discretization over the radial grid in use is adequate. Finally, the interaction
between the identified eigenmodes and the alpha-particle population is evaluated
with the  drift-kinetic code CASTOR-K, in order to assess their growth rates and
hence their linear stability. The described approach enables one to single out
the most unstable eigenmodes in a given scenario, which can then be handled with
more specialized tools. This selection capability eases the task of evaluating
alpha-particle interactions with \alfv{} eigenmodes, either for ITER scenarios
or for any other kind of scenario planning.
\end{abstract}

\section{Introduction.}

Plasma heating in the ITER burning regime will be provided mainly by an
isotropic population of fusion-born alpha particles, with thermal energy
significantly higher than that of the bulk plasma species~\cite{fasoli.2007}.
These supra thermal particles must be kept confined in the core, in order to
keep the plasma burning and prevent the vessel's walls from sustaining serious
damages. However, they can drive \alfv{} eigenmodes (AEs) unstable, which may
then enhance their transport away from the core and direct them against the
walls~\cite{breizman.2011,sharapov.2013,nabais.2012}. Therefore, to predict
alpha-particle losses and wall loads, one has to develop efficient techniques to
assess the linear stability of  all relevant modes for a given plasma
equilibrium configuration.

When modeling experimental results, measured data usually provide helpful
guidance about which AEs are interacting with the fast-particle population.
Present day diagnostics might provide the mode frequency, its radial location,
and even its structure~\cite{sharapov.2013}, narrowing considerably the number
of modes that need to be taken into account. However, no such guidance is
available when dealing with scenarios for which no experiments have been
performed yet. Therefore, the number of modes that need to be taken into
account, and have their interaction with the fast-particle population suitably
evaluated, becomes quite large. In the following, a systematic approach is
described to assess the linear stability of all AEs that can be present in a
prescribed ITER-like equilibrium, within reasonable bounds for their frequencies
and for their toroidal and poloidal mode numbers. Because the number of modes to
be handled is potentially quite large, care must be exercised when choosing the
numerical tools to work with, which must be fast and
computationally efficient.

\section{The proposed workflow.}

Given a magnetic equilibrium configuration, its corresponding AEs are first
found with the ideal-MHD code MISHKA~\cite{mihailovskii.1997}, which is used to
intensively scan a predefined range of normalized mode frequencies ($0.01 \leq
\omega/\omega_\text{A} \leq 2$, where $\omega_\text{A} = V_\text{A} / R_0$,
$R_0$ is the tokamak's major axis, and $V_\text{A}$ is the on-axis \alfv{}
velocity). For each toroidal mode number in the range $1 \leq n \leq 100$, the
frequency-scan procedure is as follows: First, the normalized-frequency interval
is sampled with a constant step size $\Delta(\omega/\omega_\text{A}) = 2 \times
10^{-5}$; Then, each frequency sample is used to provide MISHKA with a starting
guess for the mode's eigenvalue, while keeping a solution \textit{ansatz} with
$17$ poloidal harmonics ($n-1 \leq m \leq n+16$); If the ensuing eigenvalue
problem does not converge in a few iterations (in this case $7$), the resulting
solution is discarded and the next frequency sample is tried; Otherwise, the
converged eigenmode is tested to find if its discretization over the radial grid
in use ($\Delta s = 1/501$, with $s^2=\psi/\psi_\text{b}$, $\psi$ the
poloidal-field flux, and $\psi_\text{b}$ its value at the boundary) is adequate;
This procedure automatically excludes spurious, numerical-noise dominated
solutions, as well as solutions with radial wavelengths shorter than that of the
mesh size, and which cannot be adequately represented in the grid; In addition,
solutions are also discarded if their frequency is found to match the \alfv{}
continuum somewhere within the plasma, since in such circumstances they are
expected to undergo strong damping.

The frequency-scan phase ends up with a (possibly large) collection of AEs whose
interaction with the fusion-born alpha population needs to be evaluated.  This
task is carried out with the drift-kinetic code CASTOR-K~\cite{borba.1999},
which computes the energy transfer between a MHD perturbation of a given
magnetic equilibrium and a population of particles described by some equilibrium
distribution function. For every AE in the collection, CASTOR-K is run four
times: first, to compute the drive due to fusion-born alpha particles described
by a slowing-down energy distribution, and then once for each of the three
thermal species (DT-ion mixture, electrons, and Helium ash). No radiative
damping was considered in this analysis and the overall growth rate is taken to
be the sum of all four contributions.

The sizable computational effort involved in the proposed scheme is essentially
due to the very large number of AEs that need to be found by MISHKA and then
processed by CASTOR-K, and not to individual instances of either code.
Therefore, it falls into the category of trivially (or ``embarrassingly'')
parallel algorithms and a full parallelization of the proposed workflow was
implemented in the HELIOS supercomputer~\cite{helios}, taking advantage of the
independence of each AE processing. A typical workflow session in HELIOS, which
involves finding and processing a few thousand AEs, takes about $12$ to
$15$ hours if sufficient resources are available to allocate one AE to each
computing node.

\section{Results.}

\begin{figure}
\begin{center}
\includegraphics{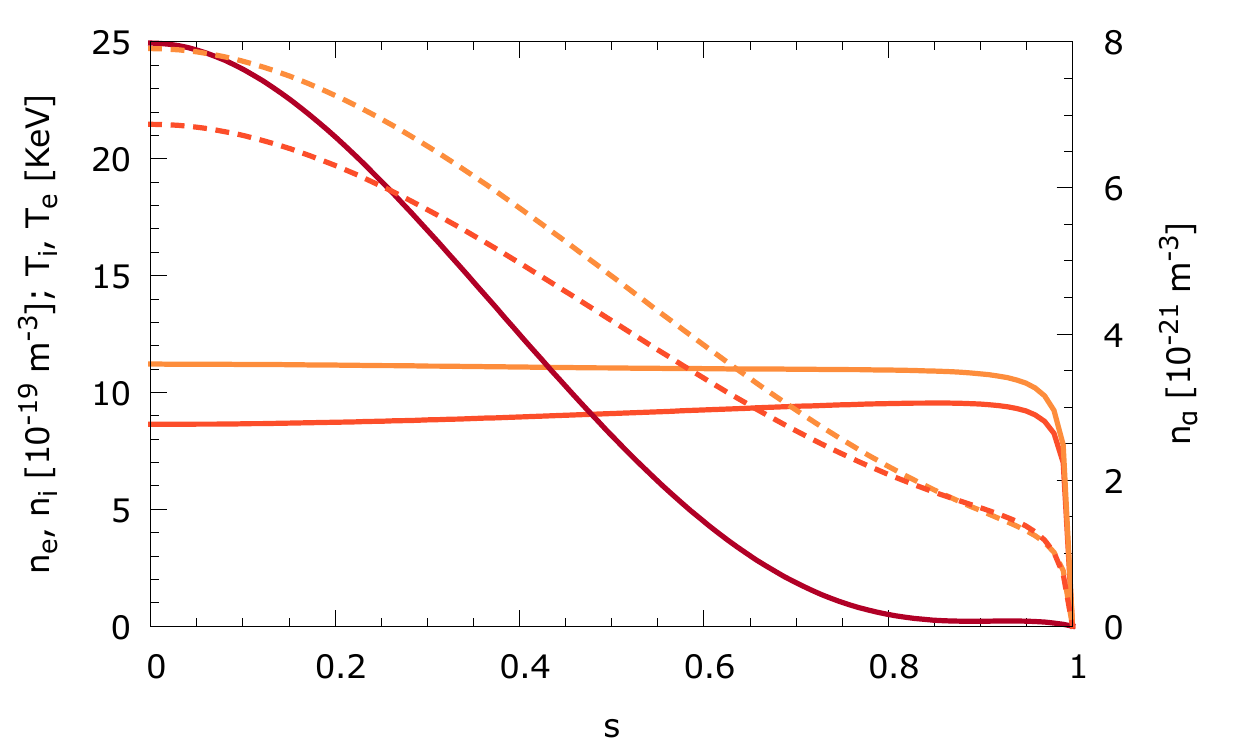}
\end{center}
\caption{
\label{fig:astra.data}
Radial distributions of alpha-particles (solid red line), DT ion (dark orange)
and electron (light orange) densities (solid lines), and temperatures (dashed
lines).}
\end{figure}

The results presented and discussed below were obtained for the ITER baseline
scenario~\cite{polevoi.2002}, with total plasma current $I_\text{p} = 15$ MA,
low shear safety-factor in the core region and $q(0)=0.987$, peaked temperature
profiles with $T_\text{i}(0)=21.5$ KeV for the DT ions and $T_\text{e}(0)=24.7$
KeV for the electrons, and flat density profiles (almost up to the edge) with
$n_\text{e}(0)=11 \times 10^{-19} \: \text{m}^{-3}$ and $n_\text{i}(0)=8.6
\times 10^{-19} \: \text{m}^{-3}$ (FIG.~\ref{fig:astra.data}). Fusion-born
alpha-particles are well confined in the core region ($s \lesssim 0.5$), with
$n_\alpha(0) = 7.9 \times 10^{-21} \: \text{m}^{-3}$ and the highest gradient of
their pressure contribution taking place at $s \approx 0.42$.

\begin{figure}
\begin{center}
\includegraphics{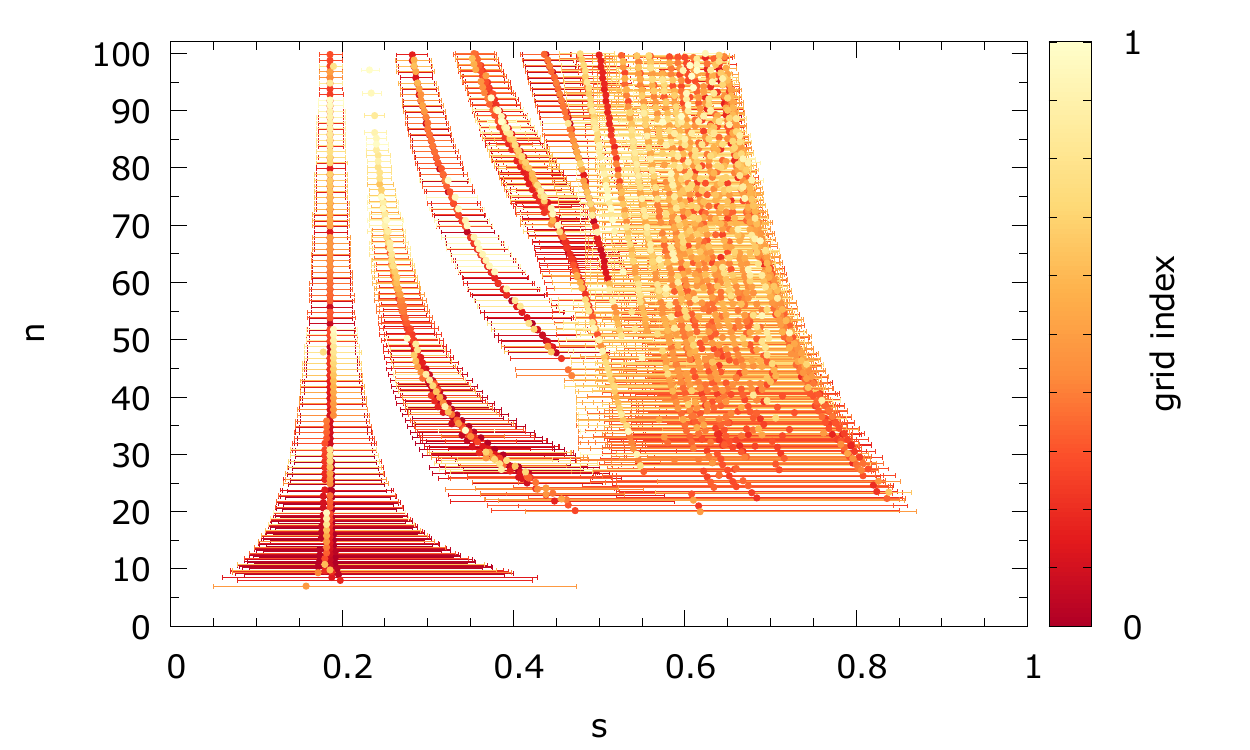}
\end{center}
\caption{
\label{fig:grid.index}
AEs radial localization and width for each toroidal number $n$, colored by their
respective grid index.}
\end{figure}

The frequency-scan procedure returns approximately $2300$ distinct AEs, which
are depicted in FIG.~\ref{fig:grid.index}. There, the dots mark the radial
location of each mode's highest peak and the lines mark its radial width. The
color denotes the grid index, which indicates the increasing difficulty of a
uniform radial grid with $501$ points to describe modes with increasingly
short-scale features. AEs with grid index greater than $1$ were excluded from
this plot and from further processing.

The results plotted in FIG.~\ref{fig:grid.index} agree, to a great extent, with
analytical estimates of mode localization and width~\cite{sharapov.2014}.
Notable exceptions are the near-vertical structure at $s \approx 0.2$ and the
structures ending at $s \approx 0.275$ and $s \approx 0.45$ for $n \approx 100$,
which correspond to solutions outside the Toroidal Alfv\'{e}n Eigenmode (TAE)
frequency gap. Another difference is the absence of modes in the left and right
upper corners, which is caused by having the AE's \textit{ansatz} limited to
$17$ poloidal harmonics. Other structures present in
reference~\cite{sharapov.2014} are not displayed in FIG.~\ref{fig:grid.index}
because the grid index of their AEs exceeds the maximum value allowed. On the
other hand, the absence of AEs for low toroidal mode numbers [$n \lesssim 20$
for TAEs and $n \lesssim 10$ for Ellipticity Alfv\'{e}n Eigenmodes (EAEs)] is
due to the particular density profile: being flat almost up to the edge, the
density profile in FIG.~\ref{fig:astra.data} causes $V_\text{A}$ (and hence the
\alfv{}-continuum frequency $\omega = k_\parallel V_\text{A}$) to drop
considerably in that region ($s \gtrsim 0.8$). This effectively closes the outer
end of the frequency gap and removes possible modes, located or extending into
there, by continuum damping.
 
\begin{figure}[t!]
\begin{center}
\includegraphics{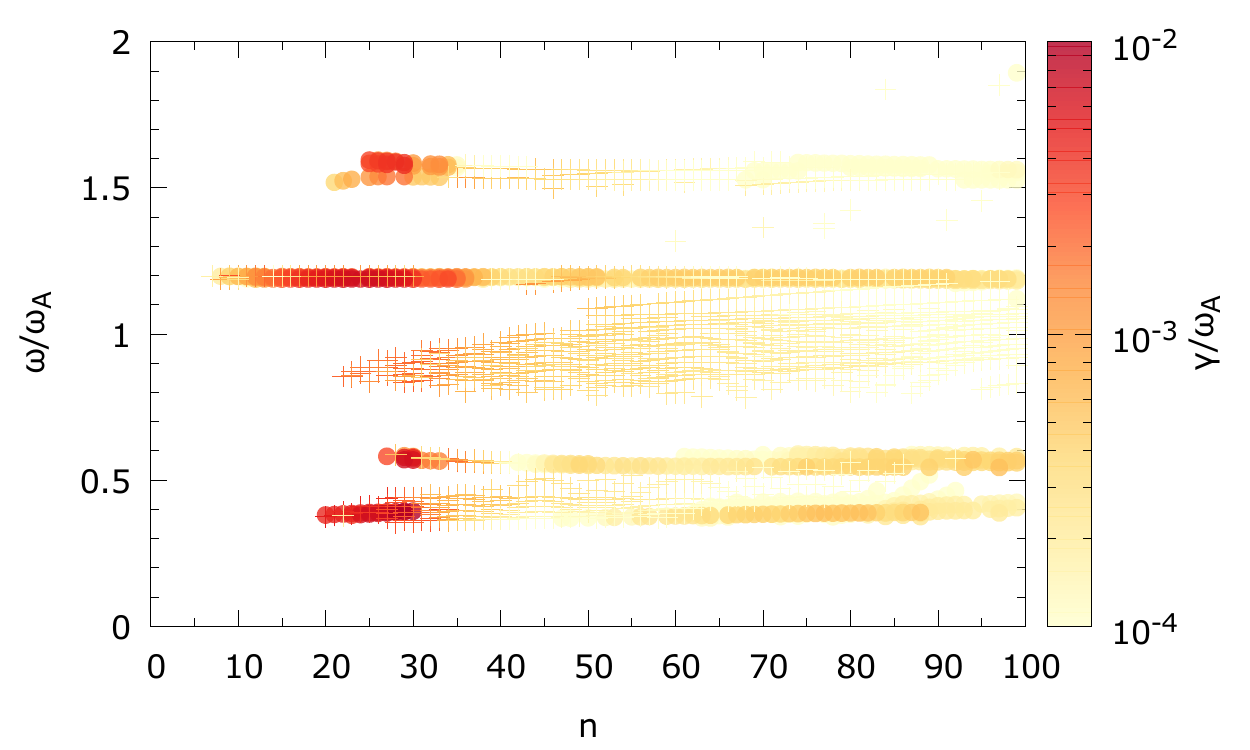}
\end{center}
\caption{
\label{fig:frequency.n}
AEs frequency distribution by toroidal number $n$, colored by their respective
growth (dots) or damping (crosses) rates.}
\end{figure}

In the next step, the linear stability of all AEs in the collection is assessed
with the CASTOR-K code. For almost all AEs, the energy exchange is dominated by
alpha particles or DT ions (energetic particles from NBI or ICRH were not
considered), whereas electrons or Helium ash have only negligible contributions.
Three frequency gaps are clearly visible in FIG.~\ref{fig:frequency.n}, each
with two branches (top and bottom): TAEs for $\omega/\omega_\text{A} \approx
0.5$, EAEs for $\omega/\omega_\text{A} \approx 1$, and Noncircular Alfv\'{e}n
Eigenmodes (NAEs) for $\omega/\omega_\text{A} \approx 1.5$. In all these three
gaps, the largest growth rates are limited to the range $20 \lesssim n \lesssim
30$. Again, there are no TAEs (stable or not) up to $n \approx 20$.  However,
unstable EAEs in the top branch start at $n \approx 10$, while all bottom-branch
EAEs are stable to alpha drive.
\begin{figure}[b!]
\begin{center}
\includegraphics{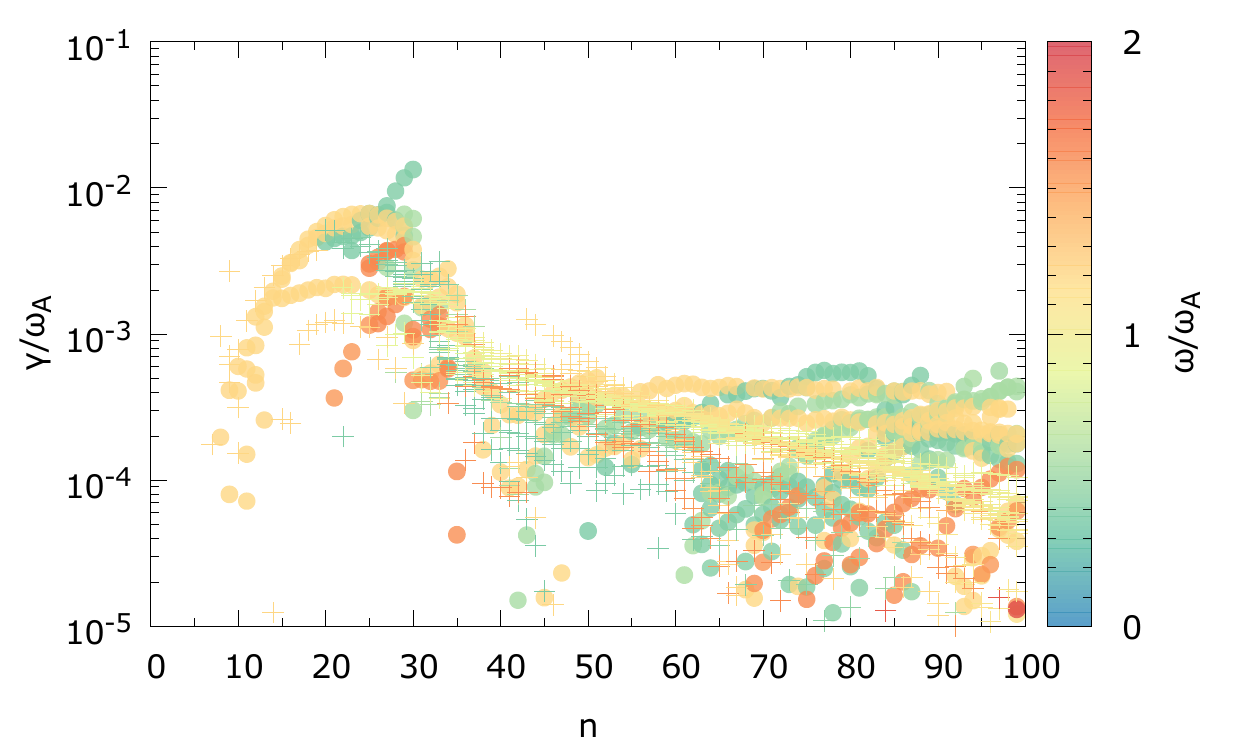}
\end{center}
\caption{
\label{fig:growthrate.n}
AEs growth (dots) and damping (crosses) rate distribution by toroidal number
$n$, colored by their respective normalized frequency.}
\end{figure}

The plot in FIG.~\ref{fig:growthrate.n} clearly shows that TAEs (greenish
symbols) are the most unstable modes, with normalized growth rates in the range
$0.5\% \lesssim \gamma/\omega_\text{A} \lesssim 2\%$ (corresponding to $1\%
\lesssim \gamma/\omega \lesssim 3.5\%$). Moreover, and in accordance with
previous estimates~\cite{sharapov.2014}, the largest growth-rate value is
attained at $n=30$. On the other hand, top-branch EAEs (light orange) and NAEs
(dark orange) have their growth rate limited to more modest ranges, these being,
respectively, $0.2\% \lesssim \gamma/\omega_\text{A} \lesssim 0.7\%$ and $0.1\%
\lesssim \gamma/\omega_\text{A} \lesssim 0.4\%$. This being said, one should
recall that radiative damping was not taken into account, which means that
computed growth rates (particularly for large $n$) may be overestimated.

\begin{figure}[b!]
\begin{center}
\includegraphics{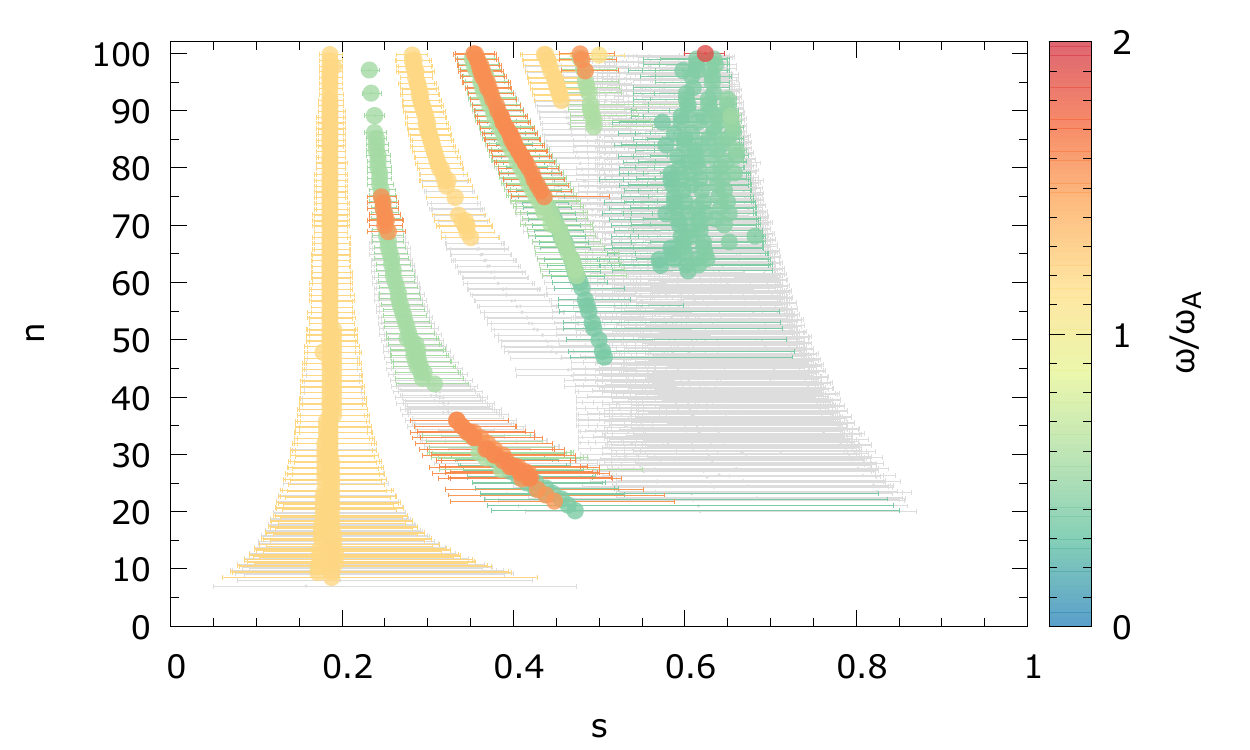}
\end{center}
\caption{
\label{fig:frequency.width}
Unstable AEs radial localization and width for each toroidal number $n$, colored
by their respective normalized frequency. Grey lines indicate stable AEs.}
\end{figure}

\begin{figure}
\begin{center}
\includegraphics{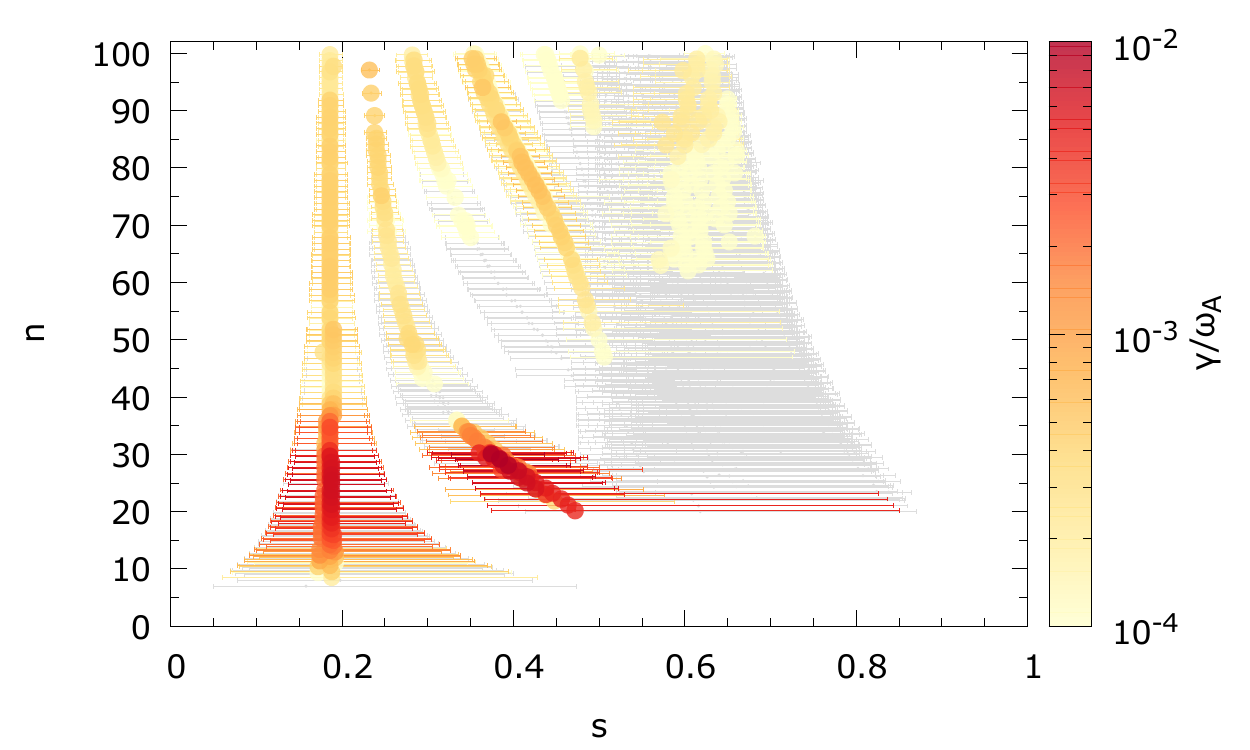}
\end{center}
\caption{
\label{fig:growthrate.width}
Unstable AEs radial localization and width for each toroidal number $n$, colored
by their respective growth rate. Grey lines indicate stable AEs.}
\end{figure}

The representation of the AEs' localization and extension can bring more insight
if it is supplemented with additional information. In
FIG.~\ref{fig:frequency.width}, unstable AEs are colored according to their
normalized frequency, enabling one to easily identify which frequency ranges
correspond to which radial structures. In particular, the near vertical
structure at $s \approx 0.2$ is seen to be made up of core-localized modes
belonging to the top branch in the EAEs' gap. Interestingly, it shows also that
the same radial structure can be shared by different frequency gaps, as is the
case of core-localized TAEs and NAEs. If unstable AEs are colored by their
growth rate instead, as in FIG.~\ref{fig:growthrate.width}, it becomes readily
obvious that the most unstable modes are short-width TAEs with $20 \lesssim n
\lesssim 30$ and localized around $0.35 \lesssim s \lesssim 0.45$.
Incidentally, one should recall from FIG.~\ref{fig:astra.data} that the highest
value of the alpha-pressure gradient takes place, precisely, at $s \approx
0.42$. The most unstable top-branch EAEs are also short-width modes in the same
toroidal mode number range, but localized deeper into the core, at $s \approx
0.2$.  Conversely, broad-width TAEs located at the outer half of the plasma ($s
\gtrsim 0.5$) are either close to marginal stability, stable, or interact with
the \alfv{} continuum, being thus excluded from the plot.


\section{Conclusions.}

A systematic approach to assess the linear stability of AEs was presented that
is particularly useful when dealing with scenarios for which no experimental
guidance is available about which particular AEs interact more intensely with
the fast particle population. In summary, it starts with an intensive scan of a
frequency and toroidal mode-number ranges to identify all possible AE
candidates, it is followed by a filtering stage to exclude non-physical
solutions, and it ends with all selected AEs having their linear stability
assessed by evaluating their energy exchange with the particle populations
present in the plasma: fusion-born alphas, DT-ion mixture, electrons, and Helium
ash. A judicious choice of the codes employed, which are fast and effective to
accomplish their intended goals, allowed the designed workflow to be efficiently
parallelized in order to take advantage of massively parallel computer systems.
Its implementation in the supercomputer HELIOS clearly indicates that the
techniques described above enable one to readily single out the most unstable
AEs for a given scenario, which can then be studied with more specialized tools.
This selection capability eases the task of evaluating
alpha-particle interactions with \alfv{} eigenmodes, either for ITER scenarios
or for any other kind of scenario planning.

Regarding the specific ITER scenario used to test the proposed approach, it was
found that most AEs with low toroidal mode number ($n \lesssim 10$) interact
with the \alfv{} continuum and that most unstable modes occur for $20 \lesssim n
\lesssim 30$, as predicted in earlier estimates~\cite{sharapov.2014}. While
broad-width TAEs located outside the core ($s \gtrsim 0.5$) were found to be
either stable, close to marginal stability, or to interact with the \alfv{}
continuum, short-width TAEs located around the maximum of the alpha-pressure
gradient ($s \approx 0.42$) were found to be the most unstable ones in all the
AE collection considered.

\section*{Acknowledgments}

All computations were carried out using the HELIOS supercomputer system at
Computational Simulation Centre of International Fusion Energy Research Centre
(IFERC-CSC), Aomori, Japan, under the Broader Approach collaboration between
Euratom and Japan, implemented by Fusion for Energy and JAEA. This work was
supported by the European Union's Horizon 2020 research and innovation programme
under the grant agreement number 633053. IST activities were also supported by
``Funda\c{c}\~{a}o para a Ci\^{e}ncia e Tecnologia'' through project
Pest-OE/SADG/LA0010/2013. The views and opinions expressed herein do not
necessarily reflect those of the European Commission.


\begin{thebibliography}{1}

\bibitem{fasoli.2007} FASOLI, A., et al., Nucl. Fusion {\bf 47} (2007) S264.

\bibitem{breizman.2011} BREIZMAN, B. and SHARAPOV, S. E., Plasma Phys. Control.
Fusion {\bf 53} (2011) 054001.

\bibitem{sharapov.2013} SHARAPOV, S. E., et al., Nucl. Fusion {\bf 53} (2013)
104022.

\bibitem{nabais.2012} NABAIS, F., et al., Nucl. Fusion {\bf 52} (2012) 083021.

\bibitem{mihailovskii.1997} MIHAILOVSKII, A., et al., Plasma Phys. Rep. {\bf 23}
(1997) 844.

\bibitem{borba.1999} BORBA, D. and KERNER, W., J. Comp. Phys. {\bf 153} (1999)
101.

\bibitem{helios} http://www.iferc.org/csc/csc.html

\bibitem{polevoi.2002} POLEVOI, A., et al., J. Fusion Res. SERIES {\bf 5 (}2002)
82.

\bibitem{sharapov.2014} SHARAPOV, S. E. and OLIVER, H., ``On \alfv{} wave
excitation in ITER baseline scenario with $I_\text{p} = 15$ MA.'', in
preparation (2014).

\end{thebibliography}
\end{document}